\documentclass[12pt]{iopart}
\usepackage{multicol,bbm}
\usepackage{graphicx,psfrag}
\usepackage{dcolumn}
\usepackage{bm}
\usepackage{amssymb}

\begin{document}

\title{Non-locality of the C- and D-theories}

\author{M. Sandstad, T. S. Koivisto and D. F. Mota}
\address{Institute of Theoretical Astrophysics, University of Oslo, 0315 Oslo, Norway}

\ead{marit.sandstad@astro.uio.no}

\begin{abstract}

Apparent similarities between non-local theories of gravity and the so-called C-theories are pointed out.
It is shown that some simple C-theories can be mapped exactly into a previously considered type of ghost-free nonlocal gravity.    
This may introduce a useful tool to tackle some infinite-order derivative theories and raises the possibility of describing 
renormalisable gravity in a new context of D-theories.
 
\end{abstract}

\submitto{\CQG}
\maketitle

  \section{Introduction}

 The success of Einsteins theory of gravity in physics has been formidable. 
 Nevertheless, in current cosmology modified theories of gravity attract considerable attention within the community as means of explaining for instance the observations from supernovae that the universe is expanding at an accelerating pace \cite{Riess_et_al1998} \cite{Perlmutter_et_al1999}. Besides the cosmological motivations to extend Einsteins General relativity at the infrared, the theory has singularities and remains unquantised at the ultraviolet; see \cite{Clifton_et_al2012},  \cite{Nojiri_et_Odintsov2011} or \cite{Capozziello_et_Laurentis2011} for recent reviews on modified theories of gravity. 

The paper \cite{Amendola_Enqvist_and_Koivisto2011} proposed a general modified gravity framework dubbed C-theories because of their definition in terms of a conformal relation of the space-time connection to the metric. This framework provides, amongst other things, a smooth transition between the metric and Palatini variational principles\cite{Koivisto032011}. For some applications of the Palatini variational principle to modified gravities and physical  implications see \cite{Olmo2011,Koivisto:2005yc,Koivisto:2006ie,pal1,pal2,pal3,Koivisto:2010jj,Koivisto:2007sq,pal4}.
The post-Newtonian (PPN) parameters were calculated in \cite{Koivisto092011} and cosmological implications for the late time universe of this theory are considered in \cite{Finke2012,Koivisto2013}.
For local metric theories, the $f(R)$ class is well known to be a special case of ghostless  higher order theories, whereas for the local Palatini theories, it was recently found \cite{Koivisto:2013kwa} that the hybrid $f(X)$ class of models 
\cite{Harko:2011nh} seems in a similar way unique in avoiding the higher order pathologies. 

However, there may exist classes of non-local, i.e. infinite-order derivative theories of gravity that are ghost-free and furthermore asymptotically free and thus renormalisable \cite{Biswas_et_al2012b}, see also \cite{Modesto:2011kw,Barvinsky:2012ts,Biswas_et_al2012c,Briscese_et_al2012}.
The particular conformal class of nonlocal models has been studied earlier \cite{Gottlober_et_al1990}, \cite{Schmidt1990} and shown to exhibit non-singular bouncing cosmological 
solutions \cite{Biswas_Mazumdar_and_Siegel}. These are defined solely in terms of the Ricci scalar and its covariant derivatives as in Eq.(\ref{eq:NonlocalAction}). They can be rewritten in terms of an infinite number of scalar fields and brought into the form of Einsteins theory by conformal transformations, which is why we refer to them as conformal nonlocal models here.  For other non-local theories, see for instance \cite{Calcagni:2010ab} and references therein. For reviews on bouncing cosmologies and other related alternatives to inflation see \cite{Novello_et_Bergliaffa2008,Brandenberger2012a,Brandenberger2012c}.
The cosmological implications of conformal nonlocal theories have been studied with considerable effort \cite{Biswas_et_al2007,Stephens2009,Biswas_et_al2010,Koshelev_and_Vernov2012,Biswas_et_al2012a,Dimitrijevic_et_al2012,Dimitrijevic_et_al2013,Koshelev2013}. To highlight the latest word on the subject,  in \cite{Biswas_et_Mazumdar2013} it was claimed that these models could explain the low multipole results from Planck \cite{PlanckXVI2013}. A related \cite{Koivisto102009} non-analytic form of the theories on the other hand has been applied for late-time cosmological modification of gravity, see e.g. \cite{Deser_et_Woodard2007,Nojiri:2007uq,Koivisto032008,Koivisto:2008dh,Capozziello:2008gu,Park:2012cp}.


In this article we consider the connection between the C-theories and the conformal nonlocal modifications of gravity. We show that for certain subsets the theories have direct mappings between each other, proving them to be dynamically equivalent. We hope that these insights may prove useful to the further investigations of these theories. 

\section{Definitions of the theories in question}

\subsection{Non-local gravity}
The non-local gravity models described in \cite{Biswas_Mazumdar_and_Siegel} can be defined by an action on the form
 \begin{equation}\label{eq:NonlocalAction}
   S_{\mathrm{nonlocal}} = \int d^d x\sqrt{-g}\left[R + R\mathcal{F}\left(\Box\right)R \right]\,,
 \end{equation}
 where $\mathcal{F}\left(\Box\right) = \sum_{n \geq{0}} f_n\Box^{n}$ is an analytic function of the d'Alembertian operator and $d$ is the dimension of space-time. The series should not truncate at any finite $n$. Heuristically one could think of the series as
 a Taylor expansion of a function of the curvature: so gravitational interactions at a space-time point are not determined solely by the curvature at that point but rather the curvature field extending over a finite region.

\subsection{C-theory}\label{sec:Definition}
The C-theory framework introduced in \cite{Amendola_Enqvist_and_Koivisto2011} is defined by use of a conformal factor $\mathcal{C}$ which is a function of the Ricci curvature scalar $\mathcal{R}$. That is we have two metrics, an unhatted metric responsible for the space-time geometry and a hatted one that generates the space-time connection, conformally related in the following way:
  \begin{equation}\label{eq:hattedUnhattedConformalRelation}
    \hat{g}_{\mu\nu} = \mathcal{C}(\mathcal{R})g_{\mu\nu} \quad \mbox{where} \quad   \mathcal{R} = g^{\mu\nu} \hat{R}_{\mu\nu}\,,
  \end{equation}
 $\hat{R}_{\mu\nu}$ being the Ricci tensor for the hatted metric. Using  (\ref{eq:hattedUnhattedConformalRelation}) we can get expressions for the hatted quantities in terms of the unhatted ones and $\mathcal{C}$. However, since $\mathcal{C}$ depends on $\mathcal{R}$ which again depends on $\mathcal{C}$ it is in general an endless recursive process to get down to just unhatted quantities. Including $\mathcal{C}$ the process is straightforward:
  \begin{equation}\label{eq:ReccursiveR}
    \mathcal{R} = R - \frac{\left(d-1\right)}{4\mathcal{C}^2}\left[4\mathcal{C}\Box\mathcal{C} + \left(d-6\right)\left(\partial\mathcal{C}\right)^2\right]\,.
  \end{equation}
In analogy with normal $f(R)$ theories the action of the full C-theory is defined as 
\begin{equation}
    S_{\mathrm{Ctheory}} = \int d^dx\sqrt{-g}f(\mathcal{R})\,.
\end{equation}


\section{The connection between the two-metric C-theory and non-local gravity}\label{sec:Nonlocal}
 
 As the expression determining $\mathcal{R}$ (\ref{eq:ReccursiveR}) is a recursive relation, that is it is a function of $\mathcal{C}$ which again is a function of $\mathcal{R}$, the expression is in general given as an infinite series of differential operators, just like a non-local gravity. However, though this indicates a certain non-locality of any generic C-theory, the exact non-locality does not necessarily correspond to that of the conformal non-local models as defined by \cite{Biswas_Mazumdar_and_Siegel}.

  Comparing the nonlocal action given in  (\ref{eq:NonlocalAction}) to our C-theory we realise that the latter is a non-local gravity of this type if we can rewrite $f(\mathcal{R})$ where $\mathcal{R}$ is given by  (\ref{eq:ReccursiveR}) into something like $R + R\mathcal{F}\left(\Box\right)R $. Rewriting  (\ref{eq:ReccursiveR}) to show the derivatives of $\mathcal{R}$ explicitly we get:
 \begin{equation}\label{eq:RandRForNonlocalConnection}
   \mathcal{R} = R - \left(d - 1\right)\left[\left(\ln\mathcal{C}\right)'\Box\mathcal{R} + h(\mathcal{C},d)\left(\partial \mathcal{R}\right)^2\right]
 \end{equation}
 where $'$ denotes derivatives with respect to $\mathcal{R}$ and $h(\mathcal{C},d)$ is a function of $\mathcal{C}$ and its derivatives and the dimensionality of the space. Its exact form is:
 \begin{equation}\label{eq:hFunction}
   h(\mathcal{C},d) = \left(\left(\ln\mathcal{C}\right)'' + \frac{d - 2}{4}\left(\left(\ln \mathcal{C}\right)'\right)^2\right)
 \end{equation}
 Having the expression in this form (\ref{eq:RandRForNonlocalConnection}) also facilitates our understanding of how the dimensionality of the theory comes into play. Whereas in the original expression (\ref{eq:ReccursiveR}) it seemed as if $d = 6$ was a special dimension, we now discover that this depends on the functional form of $\mathcal{C}$. $d=1$ is of course still a special dimension, where the C-theory always trivialises and $\mathcal{R} = R$. 
 
 To be able to use partial integration to get an expression involving only d'Alembertians and not first order derivatives, we need the function $h(\mathcal{C},d)$ given in  (\ref{eq:hFunction}) to be a constant with respect to variations in $\mathcal{R}$. Otherwise, no matter how many times we partially integrate, a term proportional to $\left(\partial \mathcal{R}\right)^2$ will always remain. One way of doing this, which will also make the rest of the steps much easier, is to assume that $\left(\ln\mathcal{C}\right)'$ is a constant, i.e. that $\mathcal{C} \propto e^{\alpha\mathcal{R}}$, where $\alpha$ is a constant. Eq. (\ref{eq:RandRForNonlocalConnection}) then becomes:
 \begin{equation}\label{eq:calRITOcalR}
   \mathcal{R} = R - \left(d - 1\right)\left[\alpha\Box\mathcal{R} + \frac{d - 2}{4}\alpha^2\mathcal{R}\Box\mathcal{R}\right]\,.
 \end{equation}
 We realise that if we keep substituting for $\mathcal{R}$ with this expression into itself only one term will remain quadratic in the curvature, while all the other terms will involve higher orders of $R$ in addition to different combinations of derivatives, and their coefficients will always be proportional to $\alpha\left(d-1\right)\left(d-2\right)$. What this observation means in terms of a more general theory is that the function $h(\mathcal{C},d)$ given in  (\ref{eq:hFunction}) does not only need to be constant, it has to be $0$ in the dimension $d$ that we are looking at. This also means that no partial integration has to be performed on the Lagrangian to proceed, the possible equality with the quadratic and conformal nonlocal model is in fact exact at the level of the Lagrangian, or not present at all. 

In (\ref{eq:calRITOcalR}) we see that it is the two-dimensional case that stands out. Getting a closed expression is then not very difficult:
 \begin{eqnarray}
   \mathcal{R} &=& R -\alpha\Box\left[R - \alpha\Box\left(R - \alpha\Box\ldots\right)\right] \nonumber\\
   &=& \sum_{n\geq 0}\left(-1\right)^n\alpha^n\Box^n R = \frac{1}{1 + \alpha\Box}R
 \end{eqnarray}
 
 For dimensions different from $d = 2$, we make an ansatz on the form $\mathcal{C} \propto \left(1 + A\mathcal{R}\right)^x$. For the function $h$ in (\ref{eq:hFunction}) to be zero in this case, we see that $x$ must take the value $\frac{4}{d-2}$, and that $A$ can stay arbitrary. Then the infinite series for $\mathcal{R}$ becomes:
 \begin{eqnarray}
   \mathcal{R} &=& R -\beta\Box\left[R - \beta\Box\left(R - \beta\Box\ldots\right)\right] \nonumber\\
   &&= \sum_{n\geq 0}\left(-1\right)^n\beta^n\Box^n R = \frac{1}{1 + \beta\Box}R
 \end{eqnarray}
 where $\beta = 4A\left(d-1\right)/\left(d-2\right)$ is an arbitrary constant that we can tune by choosing a suitable value for $A$. So the series has exactly the same functional form as in the two-dimensional case.

 From general theories of differential equations it is easy to see that these are all the C-theories that could have equivalent non-local gravity partners of the kind described in \cite{Biswas_Mazumdar_and_Siegel}. What is then needed in order to see what theories we get is the form of $f$. 

First we observe that in these cases if we put $f(\mathcal{R}) = \mathcal{R}$, or indeed make any other linear assumption for $f$, the theory seems to trivialise since the term is $R$ plus a sum of total derivative terms. However, since the total derivative is formed via an infinite series it is unclear what the exact implications for the theory will be. 

If we instead let $f(\mathcal{R})$ take the form $\mathcal{R} + c\mathcal{R}^2$, what we obtain is a theory with a term $R + cR\left(1 + \beta\Box\right)^{-2}R$, plus a total derivative term. Modulo the total derivative term this is in fact a theory of the kind described in \cite{Biswas_Mazumdar_and_Siegel} with the coefficients given by $c_n = c\left(n+1\right)\left(-\beta\right)^n$. We also realise that if we let $|\beta| > 1$, though it is unclear whether the summation is valid in that case, the theory seems to be an inverse differential operator of the kind described in \cite{Deser_et_Woodard2007}, but with an added regulator of the type described in \cite{Wetterich1998}. 

If we let $f(R)$ take higher than quadratic order, it is no longer possible to have one $i(\Box)$ operator acting only on one Ricci scalar $R$ as we will always end up with products of $\left(i(\Box)R\right)\left(j(\Box)R\right)$. Regardless of how we perform partial integrations these types of terms will be present. Therefore only the quadratic type $f$s map into conformal non-local gravities of the same kind as the ones in \cite{Biswas_Mazumdar_and_Siegel}.

\subsection{On the specific model}

To summarise our findings, in $d=4$ the C-theory specified by the Lagrangian and the conformal relation of the connection, 
\begin{equation}
f=\mathcal{R}+c\mathcal{R}^2 \quad  \mbox{and} \quad \mathcal{C} \propto \left(1 + \frac{\beta}{6}\mathcal{R}\right)^2\,, \label{c_e}
\end{equation}
respectively, is equivalent to the nonlocal model of the class (\ref{eq:NonlocalAction}) defined by the operator
\begin{equation}
\mathcal{F} = c\sum_{n\geq 0}(n+1)\left(-\beta\right)^n\Box^n\,.  \label{nl_e}
\end{equation}
A complete formalism to obtain the propagator for an arbitrary metric theory in flat space has been presented and applied to various cases in Refs.\cite{Biswas_et_al2012b,Biswas_et_al2012c,Koivisto:2013kwa}, and here we shall only state the result of such analysis for the case at hand.
We find that there are two poles. The masses and the residues associated to them are given by
\begin{eqnarray}
m^2_\pm & = & \frac{\pm \sqrt{3\left(3c+4\beta\right)}-3c-2\beta}{2\beta^2}\,,  \\
r_\pm & = & \pm \frac{c}{2}\left(c^2+\frac{4\beta c}{3}\right)^{-\frac{1}{2}}\,,
\end{eqnarray}
respectively. Obviously, one of the new propagating degrees of freedom is a ghost. The exception is the degenerate case that $\beta=-3c/4$ and the poles coincide and the system behaves like the Pais-Uhlenbeck oscillator \cite{Mannheim:2006rd,Jimenez:2012ak}. Then the mass associated to the double pole with a vanishing residue is
\begin{equation}
m = \frac{2}{\sqrt{-3c}}\,.  
\end{equation}
So given $c<0$ the model is stable. 

 \section{Conclusion and outlook}
 In this communication we noted the apparent similarities between C-theories and nonlocal gravity. In particular, we showed that there is a unique one-parameter class of C-theory models that maps exactly into a ghost-free quadratic conformal nonlocal  
 gravity model of type described for example in  \cite{Biswas_Mazumdar_and_Siegel}. More generically C-theories could be described as non-quadratic gravity with an infinite-order derivative structure that characterises non-local theories.

Thus in the specific C-theory one-to-one case studied here and more qualitatively in the generic case, the study of C-theories and conformal non-local theories of gravity may draw insights from each other. The field of conformal non-local gravity is both more vast and mature than that of C-theory and contains many theoretical insights and exact solutions especially in the high energy and inflationary context that C-theory may draw upon. The C-theory on the other hand comes with a handy two-scalar-field description \cite{Amendola_Enqvist_and_Koivisto2011}, which may be more convenient for numerical investigation also in the cases where no exact solutions can be found, as has been done in \cite{Finke2012,Koivisto2013}. 

Though such conformal theories of the form $\mathcal{F}( R,\Box )$ can introduce new interesting phenomenology without ghosts, they cannot fully address the ultraviolet problems of Einstein's theory. One needs to include Weyl-type terms in the action in order to modify the graviton propagator, which then allows to construct potentially renormalisable theories, see e.g. \cite{Biswas_et_al2012c}.  In order to generalise our mapping into such less specific actions, we need to take into account also a {\it disformal} \cite{Bekenstein:1992pj,Koivisto:2012za} contribution to the relation between the two metrics underlying the spacetime structure, such as
 \begin{equation}  \nonumber
   \hat{g}_{\mu\nu} = \mathcal{C} g_{\mu\nu} + \mathcal{D}\hat{R}_{\mu\nu}\,,
\end{equation}
where the functions $\mathcal{C}$ and $\mathcal{D}$ can depend on general curvature terms like $\mathcal{R}$ and $\hat{R}_{\mu\nu}\hat{R}^{\mu\nu}$. The connections between such D-theories and nonlocal renormalisable gravity remain to be explored.  


After the results from the Planck satellite \cite{PlanckXVI2013}, there have been suggestions that inflation might not be the most favoured scenario for the early universe \cite{Biswas_et_Mazumdar2013,Ijjas_et_al2013,Lehners_et_Steinhardt2013,Liu:2013kea}, and that cyclic or bouncing models may be favoured. Non-singular, ultraviolet complete gravity theories have predicted bouncing cosmologies, which could now also provide an explanation for the low power and odd correlations in the low multipoles observed by Planck. This certainly does not make investigations into non-local gravity related theories less interesting in the future.

\ack
D.F.M. thanks the Research Council of Norway FRINAT grant 197251/V30.
D.F.M. is also partially supported by project CERN/FP/123618/2011 and CERN/FP/123615/2011.
T.S.K. is also supported by the Research Council of Norway.

\bigskip 
\bigskip

  \bibliographystyle{unsrt}
  \bibliography{sources}

\end{document}